\documentclass[journal,draftcls,onecolumn,12pt,twoside]{IEEEtranTCOM}

\normalsize

\usepackage[nolist,nohyperlinks]{acronym}
\usepackage[T1]{fontenc}
\usepackage{cite}
\ifCLASSINFOpdf
  \usepackage[pdftex]{graphicx}
\else
  \usepackage[dvips]{graphicx}
\fi

\usepackage{amsmath}
\usepackage{subfigure}
\usepackage[cmintegrals]{newtxmath}
\usepackage{bm}
\usepackage{array}

\usepackage[table]{xcolor}
\usepackage{scrextend}

\definecolor{cadmiumgreen}{rgb}{0.0, 0.42, 0.24}
\definecolor{burntorange}{rgb}{0.8, 0.33, 0.0}

\newcolumntype{C}[1]{>{\centering\arraybackslash}p{#1}}

\begin{acronym}
\acro{OFDM}{orthogonal frequency division multiplexing}
\acro{DFT}{discrete Fourier transform}
\acro{IDFT}{inverse discrete Fourier transform}
\acro{PAM}{pulse amplitude modulation}
\acro{PSD}{power spectral density}
\acro{FFT}{fast Fourier transform}
\acro{ICC}{interference cancellation carriers}
\acro{PLC}{power line communications}
\acro{LAN}{local area network}
\acro{DMT}{discrete multitone}
\acro{CR}{cognitive radio}
\acro{HF}{high frequency}
\acro{ITU}{International Telecommunication Union}
\acro{OOBE}{out-of-band emission}
\acro{EMC}{electromagnetic compatibility}
\acro{AIC}{active interference cancellation}
\acro{CC}{cancellation carriers}
\acro{BER}{bit error rate}
\acro{LSQI}{least squares with quadratic inequality constraint}
\acro{AST}{adaptive symbol transition}
\acro{RC}{raised cosine}
\acro{RHS}{right hand side}
\acro{PAPR}{peak-to-average power ratio}
\acro{BER}{bit error rate}
\acro{NSS}{near-sidelobe-suppression}
\acro{ASS}{all-sidelobe-suppression}
\end{acronym}

\begin{document}
\title{A Generalized Spectral Shaping Method for OFDM Signals}

\author{Luis D{\'{\i}}ez, Jos{\'e} A. Cort{\'e}s, Francisco J. Ca{\~n}ete, Eduardo Martos and Salvador Iranzo}

%
%


\maketitle

\begin{abstract}

Orthogonal frequency division multiplexing (OFDM) signals with rectangularly windowed pulses exhibit low spectral confinement. Two approaches usually referred to as pulse-shaping and active interference cancellation (AIC) are classically employed to reduce the out-of-band emission (OOBE) without affecting the receiver. This paper proposes a spectral shaping method that generalizes and unifies these two strategies. To this end, the OFDM carriers are shaped with novel pulses, referred to as generalized pulses, that consist of the ones used in conventional OFDM systems plus a series of cancellation terms aimed at reducing the OOBE of the former. Hence, each generalized pulse embeds all the terms required to reduce its spectrum in the desired bands. This leads to a data-independent optimization problem that notably simplifies the implementation complexity and allows the analytical calculation of the resulting power spectral density (PSD), which in most methods found in the literature can only be estimated by means of simulations. As an example of its performance, the proposed technique allows complying with the stringent PSD mask imposed by the EN 50561-1 with a data carrier loss lower than 4\%. By contrasts, 28\% of the data carriers have to be nulled when pulse-shaping is employed in this scenario.

\end{abstract}


%
\IEEEpeerreviewmaketitle

\section{Introduction}

\IEEEPARstart{R}{ectangular} pulses conventionally employed in \ac{OFDM} yield significant \ac{OOBE}, which causes strong interference to adjacent frequency bands. This problem has acquired particular relevance with the advent of cognitive radio techniques, which allow exploiting the so-called \textit{spectrum holes} \cite{Mitola99, Haykin05}. These are legally assigned frequency bands (to a primary system) that are unused at a particular time and place and that can be employed by a secondary communication system. Reducing the \ac{OOBE} of the \ac{OFDM} signal is also of major importance in situations where spectral resources are sparse. In these circumstances, the higher the \ac{OOBE}, the larger the number of carriers that have to be nulled at both edges of the notched band and, consequently, the larger the data rate degradation. 



A plethora of methods have been proposed to reduce the \ac{OOBE} of OFDM signals (see, for instance, \cite{You14}, \cite{Chen13} and references therein). These can be classified according to the domain where they are applied, yielding time-domain and frequency-domain techniques. Filtering \cite{Faulkner00} and pulse-shaping \cite{Weinstein71} are the most popular methods of the former group.  Filtering is impractical when spectral notches are tight, as in \cite{EN50561-1}, because it requires high order filters that introduce significant distortion. Pulse-shaping has been traditionally accomplished by windowing the transmitted symbols with non-rectangular windows.  Recently, the adaptive selection of the symbol transitions and the use of a different window for each carrier have been proposed \cite{Mahmoud08}, \cite{Sahin11}. In all cases the duration of the OFDM symbols is extended, causing a data rate loss and an increment in the transmitted energy per bit. 

The simplest frequency-domain method consists in nulling carriers at the band edges. The inefficiency of this strategy motivates the \ac{AIC} technique, in which a set of carriers are used to lower the \ac{OOBE} \cite{Bingham96}. These carriers are referred to as \ac{CC} and their modulating values are a function of the ones that modulate the data carriers. Methods based on this concept have been proposed in \cite{Yamaguchi04}, \cite{Brandes06}, \cite{vandeBeek08}. A related technique, in which the cancellation signal consists of tones spaced closer than the intercarrier spacing, has been proposed in \cite{Qu10}. Since the \ac{CC} do not convey information but consume a fraction of the total power, the data rate is reduced and the transmitted energy per bit increases. Nevertheless, both drawbacks are usually unimportant because the number of required \ac{CC} is generally low. An alternative frequency-domain approach consists in applying some sort of precoding to the values that modulate the data carriers. This has yielded a large set of methods that differ in the characteristics of the employed precoding \cite{You14}, \cite{Cosovic06}, \cite{Chung08},\cite{vandeBeek09bis}, \cite{Ma11}. 

\ac{OOBE} reduction techniques can be also classified according to their sidelobe suppression range, resulting in two categories referred to as \ac{NSS} and \ac{ASS} \cite{Chen13}. While the latter achieves significant reductions of the sidebands at distant frequencies, the former are more effective to create narrow notches in the passband
. Precoding and most time-domain methods belong to the \ac{ASS} group. On the contrary, \ac{AIC} methods generally belong to the \ac{NSS} category. Due to this, the combined use of \ac{AIC} and pulse-shaping has been proposed \cite{Brandes10}. 

Interestingly, many of the aforementioned methods can be applied transparently to the receiver, which allows existing standards to achieve large \ac{OOBE} reductions while guarantying forward and backward compatibility. As an example, this is useful to make the latest release of \ac{PLC} systems based on the ITU-T Rec. G.9964 \cite{G.9964}, which was defined in 2011, to comply with the dynamic frequency exclusion mechanism stated in the EN50561-1, which was issued in 2013, with a much smaller data rate loss than by nulling carriers. \ac{AIC} strategies are particularly appropriate for this purpose, since receiver operation remains unaltered. On the contrary, precoding methods require the receiver to be modified so that it becomes aware of the precoding. Otherwise, an increment in the \ac{BER} occurs. 

This work focuses on \ac{OOBE} reduction techniques that can be applied transparently to the receiver. In this context, the following contributions are made: 
\begin{itemize}
\item It defines a general framework that unifies time-domain and \ac{AIC} strategies, which have been traditionally treated as disjoint approaches. This allows expressing methods already proposed in the literature as particular solutions of the presented one. To this end, the data carriers employ a novel pulse that is referred to as generalized pulse because it consists of the one used in conventional OFDM systems plus a cancellation term that is aimed at reducing the \ac{OOBE} of the former. The waveform of the cancellation term is computed subject to the constraint that orthogonality at the receiver must be preserved.

\item It proposes two strategies for the design of the cancellation term that result in efficient \ac{IDFT}-based implementations. 

\item The proposed method offers two main advantages over previous \ac{AIC} and time-domain ones. First, it achieves significantly larger \ac{OOBE} reductions by jointly optimizing the frequency and time-domain cancellation terms. Second, it yields a data-independent optimization process that can be accomplished offline in advance. This overcomes the problem of many proposals found in the literature, which are not used in practice because an optimization problem has to be solved for each \ac{OFDM} symbol. As a consequence, the \ac{PSD} obtained with these techniques can only be estimated by means of simulations. On the contrary, the method presented here allows to analytically calculate the \ac{PSD}. 

\end{itemize}

The rest of the paper is organized as follows. Section \ref{Notation and background} introduces the employed notation, defines some elements that will be used throughout of the paper and summarizes the background of the problem. The definition of the generalized pulse, along with its optimization procedure, and a discussion of the complexity of its associated transmitter are given in Section \ref{Generalized method for spectral shaping}. 
Section \ref{Reduced-complexity strategies} presents a catalog of design strategies that simplify the optimization problem and the transmitter implementation. Section \ref{Relation to previous methods and novelties of the proposal} highlights its relation to previous works. The performance assessment of the proposed  method is provided in Section \ref{Results}. Finally, Section \ref{Conclusion} recapitulates the main elements of the work.

\section{Notation and background}
\label{Notation and background}

\subsection{Notation and definitions}
Scalar variables are written using italic letters. Matrices and column vectors are written in boldface, the former in capital letters. Sets are denoted using calligraphic letters, e.g. $\mathcal{A}$, and their cardinality as $|\cdot|$. The Hermitian, the conjugate and transpose operators are denoted as $(\cdot)^H$, $(\cdot)^*$ and $(\cdot)^T$, respectively. The imaginary unit is written as $j=\sqrt{-1}$. The superscripts $(\cdot)^{\Re}$ and $(\cdot)^{\Im}$ denote the real and the imaginary parts of a complex value. $\mathbf{I}_M$ is the $M \times M$ identity matrix, while $\mathbf{0}_{M,N}$ is an $M \times N$ zero matrix. An $M \times M$ diagonal matrix with elements $\left\{x_1, \ldots, x_{M}\right\}$ is denoted as $\textrm{diag}\left(x_1, \ldots, x_{M}\right)$. The complex exponential will be written as $w_N^{kn}=e^{j\frac{2\pi}{N}kn}$, and the $N$-point \ac{IDFT} matrix is columnwise expressed as $\mathbf{W}_N=\left[\mathbf{w}_N^{0}, \ldots, \mathbf{w}_N^{N-1}\right]$, where $\mathbf{w}_N^k=\left[w_N^{0}, \ldots, w_N^{k(N-1)}\right]^T$. 

The considered OFDM system has $N$ carriers, which can be classified into three sets according to their functionality: data, cancellation and null. Data carriers are used for conveying information and their indexes are given by the set $\mathcal{D}=\left\{d_1, \ldots, d_{|\mathcal{D}|}\right\}$. Cancellation carriers are exclusively used to shape the spectrum and their indexes are $\mathcal{C}=\left\{c_1, \ldots, c_{|\mathcal{C}|}\right\}$. The union of both sets is denoted as $\mathcal{K} = \mathcal{D} \cup \mathcal{C}=\left\{k_1, \ldots, k_{|\mathcal{K}|}\right\}$. Finally, null carriers are those with no allocated power.  

\subsection{OFDM signal generation and \ac{PSD} expression}
\label{Power Spectral Density of an OFDM signal}
The discrete-time lowpass-equivalent of an \ac{OFDM} signal can be written as, 
\begin{equation}
x(n) = \sum_{i=-\infty}^{\infty} x_i(n-iN_s),
\label{ec.0}
\end{equation}
where $N_s=N+N_{GI}$ is the symbol period, $N_{GI}$ represents the number of samples of the guard interval and the $i$-th OFDM symbol is given by
\begin{equation}
x_i(n) = \sum_{k \in \mathcal{D}} s_k(i) p_k(n),
\label{ec.01bis}
\end{equation}
where $k$ is the carrier index, $p_k(n)$ is the pulse used in carrier $k$ and $s_k(i)$ denotes the $i$-th modulating symbol transmitted in carrier $k$.

In a rectangularly windowed OFDM system, the basic pulse $p_k(n)$ is obtained by modulating a rectangular shaping pulse, $g(n)$, 
\begin{equation}
p_k(n) = g(n)w_N^{k(n-N_{GI})},
\label{ec.1}
\end{equation}
where $g(n)$ is non-zero only in the interval $n \in \left\{0,\ldots, L-1\right\}$, with $L=N_s$. Using vector notation $p_k(n)$ can be expressed as $\mathbf{p}_k = \left[ p_k(0), \ldots, p_k(L-1) \right]^T$.

\begin{figure}[h!]
\centering
  \includegraphics[width=0.75\columnwidth]{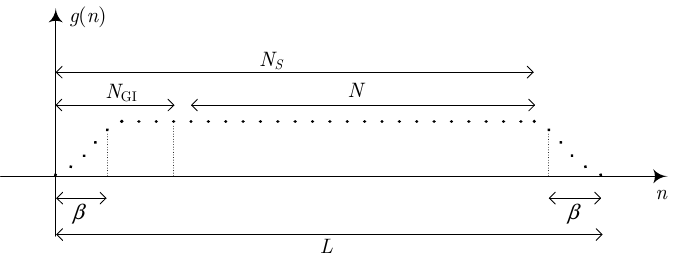}
  \caption{Waveform of $g(n)$ with smooth transitions of $\beta$ samples at both ends.} 
\label{Fig.0}
\end{figure}

Shaping pulses with smooth transitions, as the one shown in Fig. \ref{Fig.0}, are typically used to reduce the \ac{OOBE} \cite{Weinstein71}. In this case, $L=N_s+\beta$ and the waveforms of successive symbols overlap $\beta$ samples at both symbol ends. Since the receiver discards the first $N_{GI}$ samples of each symbol and computes the \ac{DFT} of the $N$ following ones, it is unaffected by this spectral shaping as long as $\beta<N_{GI}$. However, the cyclic prefix reduces to $N_{GI}-\beta$ because of the smoothed samples at the beginning of the symbol \cite{Weiss04bis}. Hence, the following condition must also hold to avoid degrading the \ac{BER} when the signal propagates through a frequency selective
channel,
\begin{equation}
L_c<N_{GI}-\beta,
\label{ec.1bis_2}
\end{equation}
where $L_c$ is the channel impulse response length.

Using matrix notation $x_i(n)$ can be expressed as

\begin{equation}
\mathbf{x}_i=\mathbf{P}_{\mathcal{D}} \mathbf{s}_{\mathcal{D}}(i),
\label{ec.1bis}
\end{equation}
where $\mathbf{P}_{\mathcal{D}}=\left[\mathbf{p}_{d_1}, \ldots, \mathbf{p}_{d_{|\mathcal{D}|}},\right]$ is an $L \times |\mathcal{D}|$ matrix defined in terms of the column vector $\mathbf{p}_k$ with $k=\left\{d_1,\cdots, d_{|\mathcal{D}|}\right\}$ and $\mathbf{s}_{\mathcal{D}}(i)=\left[s_{d_1}(i), \ldots, s_{d_{|\mathcal{D}|}}(i)\right]^T$. 

When $g(n)$ has the waveform shown in Fig. \ref{Fig.0}, expression (\ref{ec.1bis}) can be conveniently expressed as 
\begin{equation}
\mathbf{x}_i= \mathbf{P}_{\mathcal{D}} \; \mathbf{s}_{\mathcal{D}}(i) =\mathbf{G} \;
\pmb{\Delta}_{N_{GI},\beta} \; \mathbf{W}_N^{\mathcal{D}} \; \mathbf{s}_{\mathcal{D}}(i),
\label{ec.1bisbis}
\end{equation}
with $\mathbf{W}_N^{\mathcal{D}}=\left[\mathbf{w}_N^{d_1}, \ldots, \mathbf{w}_N^{d_{|\mathcal{D}|}}\right]$, $\mathbf{G}=\textrm{diag}\Big(g\left(0\right), \ldots, g\left(L-1\right)\Big)$ and where $\pmb{\Delta}_{N_{GI},\beta}$ is an $L \times N$ matrix that performs the cyclic extension of the IDFT output at both ends,
\begin{equation}
\pmb{\Delta}_{N_{GI},\beta}=\left( 
\begin{array}{cc} 
\mathbf{0}_{N_{GI},N-N_{GI}} & \mathbf{I}_{N_{GI}} \\ 
\;\;\;\;\;\;\;\;\;\;\;\;\;\;\;\; \mathbf{I}_N \\ 
\mathbf{I}_{\beta} & \mathbf{0}_{\beta,N-\beta} 
\end{array} 
\right).
\label{ec.1bisbisbis}
\end{equation}

Expression (\ref{ec.1bisbis}) is the matrix form of the \ac{IDFT}-based implementation of an OFDM transmitter. Modulating symbols are firstly passed through the \ac{IDFT}. Then, the last $N_{GI}$ samples are repeated at the beginning and the first $\beta$ samples are repeated at the end of the symbol. Finally, the first and the last $\beta$ samples are shaped to smooth the transitions between symbols. 

The \ac{PSD} of the OFDM signal\footnote{Since the OFDM signal is cyclostationary, the magnitude of interest for the \ac{OOBE} is the time-averaged \ac{PSD}, which for conciseness will be simply referred to as \ac{PSD}.} depends on the autocorrelation of the sequence of modulating symbols and on the spectrum of the shaping pulse \cite{VanWaterschoot10}. Hence, spectral shaping methods act on these elements. Frequency-domain ones introduce correlation between the transmitted symbols, either by precoding the values transmitted on the data carriers or by adding \ac{CC} whose modulating values are a function of the data ones, while time-domain techniques modify the spectrum of the shaping pulse. 

Assuming that the sequence $s_k(i)$ transmitted in each data carrier is white and that sequences transmitted in different carriers are independent, the \ac{PSD} of (\ref{ec.0}) can be expressed as
\begin{equation}
S(f) = \frac{1}{N_s}\sum_{k \in \mathcal{D}} \sigma_k^2 \left|P_k(f)\right|^2,
\label{ec.2}
\end{equation}
where $\sigma_k^2$ is the variance of $s_k(i)$, $f\in(-1/2,1/2]$ denotes the discrete-time normalized frequency and $P_k(f)$ is the Fourier transform of $p_k(n)$, which using matrix notation can be compactly written as, 
\begin{equation}
P_k(f)=\mathbf{f}_L^H \! (f) \; \textbf{p}_k,
\label{ec.3}
\end{equation}
with $\mathbf{f}_L^H \! (f) = \left[ 1,  e^{-j2 \pi f}, \ldots, e^{-j2 \pi f \left(L-1\right)} \right]$.

Let us denote the frequency band where the \ac{OOBE} has to be reduced by $\mathcal{B}$. The power of the OFDM signal in $\mathcal{B}$ is given by,
\begin{equation}
P_{\mathcal{B}}=\int_{\mathcal{B}} S(f) df = \frac{1}{N_s}\sum_{k \in \mathcal{D}} \sigma_k^2 E_{k,\mathcal{B}},
\label{ec.4}
\end{equation}
where $E_{k,\mathcal{B}}$ is the energy of the pulse transmitted in the $k$-th carrier in the frequency band $\mathcal{B}$, 
\begin{equation}
E_{k,\mathcal{B}}=\int_{\mathcal{B}} \left| P_k(f)\right|^2 df = \mathbf{p}_k^H \pmb{\Phi}_{\mathcal{B}} \; \mathbf{p}_k,
\label{ec.5}
\end{equation}
where $\pmb{\Phi}_{\mathcal{B}}$ is a $LxL$ Hermitian Toeplitz matrix that depends only on the considered frequency range, 
\begin{equation}
\pmb{\Phi}_{\mathcal{B}}=\int_{\mathcal{B}} \mathbf{f}_L(f) \; \mathbf{f}_L^H \! (f) df.
\label{ec.6}
\end{equation}

\section{Generalized method for spectral shaping}
\label{Generalized method for spectral shaping}
\subsection{Generalized pulse definition}
\label{Generalized pulse definition}
The proposed method reduces the \ac{OOBE} in the frequency band $\mathcal{B}$ by changing the set of pulses employed in (\ref{ec.1bis}) to, 
\begin{equation}
\mathbf{x}_i=\mathbf{H}_{\mathcal{D}} \mathbf{s}_{\mathcal{D}}(i),
\label{ec.6bis}
\end{equation}
with $\mathbf{H}_{\mathcal{D}}=\left[\mathbf{h}_{d_1}, \ldots, \mathbf{h}_{d_{|\mathcal{D}|}}\right]$ and where $\mathbf{h}_k$ is a novel pulse, which from now on will be referred to as generalized pulse. 

Each $\mathbf{h}_k$ is independently designed to minimize its spectrum in the notched band, $H_k(f)$ with $f \in \mathcal{B}$, while preserving orthogonality at the receiver. This is achieved by defining $\mathbf{h}_k$ as
\begin{equation}
\mathbf{h}_k = \mathbf{p}_k + \mathbf{P}_{\mathcal{C}} \, \pmb{\alpha}_k + \mathbf{t}_k, \;\; k\in \mathcal{D}.
\label{ec.7}
\end{equation}

As seen, the first term on the \ac{RHS} of (\ref{ec.7}) is the basic pulse defined in (\ref{ec.1}), which bears the information. The remaining two terms are exclusively added to reduce the spectrum of the former in the notched band. Hence, the second term on the \ac{RHS} of (\ref{ec.7}) is a linear combination of the pulses transmitted in the set of \ac{CC}, whose weights $\pmb{\alpha}_k=\left[\alpha_k(1), \ldots, \alpha_k\left(|\mathcal{C}|\right)\right]^T$ are determined by means of an optimization process. They will be referred to as \textit{cancellation pulses}. 

The third term on the \ac{RHS} of (\ref{ec.7}) is the vector form of the pulse $t_k(n)$, which will be referred to as \textit{transition pulse}. It is intended to modify the boundaries of the previous terms by acting over the first and the last $\beta$ samples. Hence, its samples are non zero-valued only in the region $n=\left\{0, \dots, \beta-1, L-\beta, \ldots, L-1 \right\}$. In order to yield a compact formulation of the problem it is expressed as $\mathbf{t}_k=\mathbf{T}\,\pmb{\zeta}_k$, where $\mathbf{T}$ is the $L \times 2\beta$ matrix 

\begin{equation}
\mathbf{T}= \left[
\begin{array}{c}
\begin{matrix}\mathbf{I}_{\beta} & \mathbf{0}_{\beta,\beta} \end{matrix}\\
\;\;\;\;\;\;\mathbf{0}_{L-2\beta,2\beta}\\
\begin{matrix}\mathbf{0}_{\beta,\beta} & \mathbf{I}_{\beta} \end{matrix}
\end{array}
\right],
\label{ec.8}
\end{equation}
and $\pmb{\zeta}_k=\left[\zeta_k(1), \ldots,\zeta_k(2\beta)\right]^T$ is determined by means of an optimization process.
 
Expression (\ref{ec.7}) can then be compactly written as,
\begin{equation}
\mathbf{h}_k = \mathbf{p}_k + \Big[\mathbf{P}_{\mathcal{C}} \;\; \mathbf{T} \Big] \left[ \begin{array}{c} \pmb{\alpha}_k \\ \pmb{\zeta}_k \end{array}\right]= \mathbf{p}_k + \pmb{\Pi} \, \pmb{\gamma}_k, \;\; k\in \mathcal{D},
\label{ec.9}
\end{equation}
where
\begin{equation}
\pmb{\Pi}=\Big[\mathbf{P}_{\mathcal{C}} \;\; \mathbf{T} \Big],\;\; \pmb{\gamma}_k=\left[ \begin{array}{c} \pmb{\alpha}_k \\ \pmb{\zeta}_k \end{array} \right].
\label{ec.9bis}
\end{equation}


It must be emphasized that the data symbols in (\ref{ec.6bis}) can be retrieved by a conventional receiver without influencing the \ac{BER}. This is because the first term in the \ac{RHS} of (\ref{ec.7}) equals the one of a conventional \ac{OFDM} and the remaining two are transparent to the receiver. This can be seen by noting that $k$-th output of the DFT computed by the receiver can be expressed as $\sum_{n=N_{GI}+1}^{N_{GI}+N}h_j(n)w^{-kn}_{N}, \; \forall j,k \in \mathcal{D}$, and that its value is unaffected by the second term in the RHS of (\ref{ec.7}) because $\sum_{n=N_{GI}+1}^{N_{GI}+N}p_c(n)w^{-kn}_{N}=0, \; \forall k \in \mathcal{D}, \forall c \in \mathcal{C}$. Similarly, since $t_k(n)=0$ for $N_{GI}+1\leq n \leq N+N_{GI}$, $\sum_{n=N_{GI}+1}^{N_{GI}+N}t_j(n)w^{-kn}_{N}=0, \; \forall j,k \in \mathcal{D}$ as long as (\ref{ec.1bis_2}) holds.

\begin{figure}[h!]
\centering
\includegraphics[width=0.6\columnwidth]{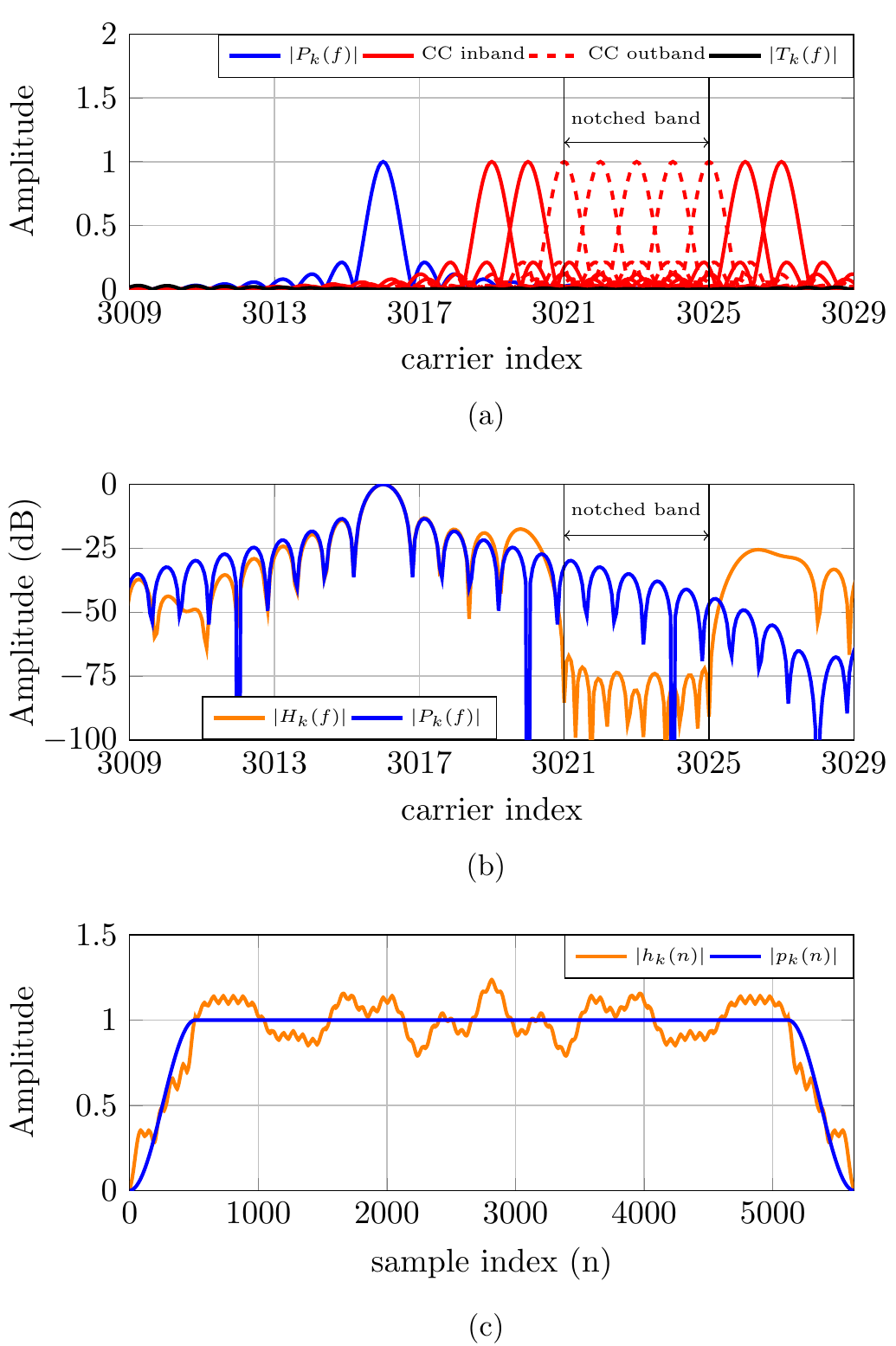}
\caption{Illustrative representation of the generalized pulses. (a) Spectrum of its constituent terms: basic pulse, \ac{CC} and transition pulse; (b) Spectrum of the basic pulse and of the generalized one;  (c) time-domain modulus of the basic pulse and of the generalized one.} 
\label{Fig.1}
\end{figure}

Fig. \ref{Fig.1} depicts the waveform and the spectrum of a generalized pulse. Fig. \ref{Fig.1} (a) shows its constituent terms in the frequency domain: the basic pulse, the \ac{CC} and the transition pulse. For illustrative purposes, it has been assumed that the \ac{OOBE} has to be reduced in the subband denoted as 'notched band' and that $g(n)$ has a \ac{RC} shape. \ac{CC} located in the notched band have been labeled  as \ac{CC} outband, while those that were previously used as data carriers are denoted as \ac{CC} inband. The latter achieve larger \ac{OOBE} suppression than the former at the cost of a data rate penalty. Fig. \ref{Fig.1} (b) plots the spectrum of the generalized pulse and of its constituent basic pulse. As seen, the spectrum of the generalized pulse in the notched band is about $35$ dB below the one of the basic pulse. Fig. \ref{Fig.1} (c) shows the modulus of $h_k(n)$ and of $p_k(n)$. For the sake of clarity, they have been depicted using lines instead of dots. The ripple in the flat region of $h_k(n)$ is due to the \ac{CC}, while its shape at the edges is mainly due to the transition pulses. 

The key feature of the proposed method is that it lowers the \ac{OOBE} by reducing the energy of the generalized pulses in the notched band, which is achieved by diminishing the magnitude of $|H_k(f)|$ as illustrated in Fig. \ref{Fig.1} (b). To this end, the basic pulses, $p_k(n)$, are modified but preserving orthogonality at the receiver. This is accomplished by adding cancellation pulses, $p_c(n)$ with $c \in \mathcal{C}$, and transition pulses that are time-limited to the boundaries of the symbol, $t_k(n)$. Their amplitudes are determined to minimize the \ac{OOBE} of the generalized pulse, $h_k(n)$, in the notched band. Since it is the pulse waveform employed in each carrier what is optimized, the process is data-independent. This contrasts with the conventional approaches, in which the OFDM is firstly generated and then its OOBE is reduced by adding a single cancellation term for the whole symbol. 

\subsection{Optimization procedure}
\label{Optimization procedure}
This subsection describes the optimization process of the generalized pulse proposed in (\ref{ec.9}). This is done with the objective of minimizing the power of the \ac{OFDM} signal in the frequency band $\mathcal{B}$, which can achieved by minimizing the energy of the generalized pulses employed in the data carriers, as shown in (\ref{ec.4}). Accordingly, each $\pmb{\gamma}_k$ is determined as
\begin{equation}
\hat{\pmb{\gamma}}_k = \arg \min_{\pmb{\gamma}_k} \left\{E_{k,\mathcal{B}}\right\},  \;\;\forall k \in \mathcal{D},
\label{ec.10bisbisbis}
\end{equation}
where
\begin{equation}
\begin{aligned}
E_{k,\mathcal{B}}&=\int_{\mathcal{B}} \left| H_k(f)\right|^2 df = \mathbf{h}_k^H \pmb{\Phi}_{\mathcal{B}} \mathbf{h}_k \\
&=\left(\mathbf{p}_k^H + \pmb{\gamma}_k^H \pmb{\Pi}^H \right) \pmb{\Phi}_{\mathcal{B}} \left(\mathbf{p}_k + \pmb{\Pi}\; \pmb{\gamma}_k \right).
\end{aligned}
\label{ec.11}
\end{equation}
Interestingly, since $\mathbf{h}_k$ generalizes $\mathbf{p}_k$, so does (\ref{ec.11}) generalizes (\ref{ec.5}), being equal when $\pmb{\gamma}_k=\mathbf{0}_{|\mathcal{C}|+2\beta,1}$.  

$E_{k,\mathcal{B}}$ is a quadratic form in the coefficients $\pmb{\gamma}_k$, hence, its minimum energy solution is given by
\begin{equation}
\hat{\pmb{\gamma}}_k=-\left( \pmb{\Pi}^H \pmb{\Phi}_{\mathcal{B}} \pmb{\Pi} \right)^{-1} \pmb{\Pi}^H \pmb{\Phi}_{\mathcal{B}} \mathbf{p}_k.
\label{ec.12}
\end{equation}

It is worth mentioning that both $\pmb{\Pi}$ and $\pmb{\Phi}_{\mathcal{B}}$ are independent of the carrier index, $k$. The former only depends on the set of employed \ac{CC} and the latter on the band where the \ac{OOBE} has to be reduced, $\mathcal{B}$.

Unfortunately, the solution in (\ref{ec.12}) may cause undesirable peaks in the passband of the \ac{PSD} for certain OFDM parameters. This can be avoided by imposing a constraint on the absolute value of the real and imaginary parts of $\pmb{\alpha}_k$ and $\pmb{\zeta}_k$,
\begin{equation}
\begin{aligned}
&
\begin{matrix}
|\Re \left\{\alpha_k(j)\right\}|, |\Im \left\{\alpha_k(j)\right\}| \leq \epsilon_{\scalebox{0.6}{CC}} \;\;\textrm{for} \; j=\left\{1, \ldots, |\mathcal{C}|\right\},
\end{matrix} \\
&
\begin{matrix}
|\Re \left\{\zeta_k(r)\right\}|, |\Im \left\{\zeta_k(r)\right\}|\leq\epsilon_{t} \;\;\textrm{for} \; r=\left\{1, \ldots, 2\beta\right\},
\end{matrix}
\end{aligned}
\label{ec.12_a}
\end{equation}
which leads to an optimization problem that can be easily solved by means of quadratic programming techniques.  

It is again interesting to highlight the differences between the proposed optimization and the classical \ac{AIC} and time-domain methods in \cite{Brandes06}, \cite{Mahmoud08}. The problem in (\ref{ec.10bisbisbis}) is solved for each data carrier and is data-independent. Its objective is to reduce the energy of each carrier in the notched band, which is achieved by modifying the employed pulse. Hence, it can be performed offline in advance. On the contrary, the objective of the problems in \cite{Brandes06}, \cite{Mahmoud08} is to lower the spectrum of each \ac{OFDM} symbol (or pair of symbols). Hence, the optimization has to be accomplished for each symbol and depends on the transmitted data. This makes a significant difference to the computational complexity of the proposed method.  

\subsection{Transmitter implementation}
\label{Generalized method for spectral shaping-Transmitter implementation}
When the generalized pulses defined in (\ref{ec.9}) are employed, the $i$th OFDM symbol can be expressed as
\begin{equation}
\mathbf{x}_i = \mathbf{H}_{\mathcal{D}} \mathbf{s}_{\mathcal{D}}(i) = \mathbf{P}_{\mathcal{D}} \, \mathbf{s}_{\mathcal{D}}(i) + \mathbf{P}_{\mathcal{C}} \mathbf{A}_{\mathcal{D}}\, \mathbf{s}_{\mathcal{D}}(i) + \mathbf{T} \, \pmb{Z}_{\mathcal{D}}\,\mathbf{s}_{\mathcal{D}}(i),
\label{ec.18}
\end{equation}
where $\mathbf{A}_{\mathcal{D}}=\left[\hat{\pmb{\alpha}}_{d_1}, \ldots, \hat{\pmb{\alpha}}_{d_{|\mathcal{D}|}}\right]$ and $\pmb{Z}_{\mathcal{D}}=\left[\hat{\pmb{\zeta}}_{d_1}, \ldots, \hat{\pmb{\zeta}}_{d_{\mathcal{D}}}\right]$.

The first and the second terms on the \ac{RHS} of expression (\ref{ec.18}) can be efficiently implemented using a single \ac{IDFT},
\begin{equation}
\mathbf{x}_i = \mathbf{G} \, \pmb{\Delta}_{N_{GI},\beta} \, \left( \mathbf{W}_N^{\mathcal{D}}\,\mathbf{s}_{\mathcal{D}}(i) + \mathbf{W}_N^{\mathcal{C}} \; \mathbf{A}_{\mathcal{D}} \, \mathbf{s}_{\mathcal{D}}(i) \right) + \mathbf{T} \, \pmb{Z}_{\mathcal{D}}\,\mathbf{s}_{\mathcal{D}}(i),
\label{ec.18bis}
\end{equation}
where $\mathbf{W}_N^{\mathcal{C}}=\left[\mathbf{w}_N^{c_1},\ldots, \mathbf{w}_N^{c_{|\mathcal{C}|}}\right]$. The additional implementation complexity of these terms with respect to the conventional OFDM transmitter is the computation of $\mathbf{A}_{\mathcal{D}} \mathbf{s}_{\mathcal{D}}(i) $, which involves the product of the $|\mathcal{C}| \times |\mathcal{D}|$ matrix $\mathbf{A}_{\mathcal{D}}$ by the $|\mathcal{D}| \times 1$ vector $\mathbf{s}_{\mathcal{D}}(i)$. The number of complex products associated to the implementation of the transition pulses is $2\beta|\mathcal{D}|$. Since $\beta$ is usually large, section \ref{Reduced-complexity strategies} proposes some particular waveforms for $t_k(n)$ that admit an efficient \ac{IDFT}-based implementation. 

\section{Design strategies with reduced implementation complexity}
\label{Reduced-complexity strategies}


\subsection{Reducing the set of data carriers that use generalized pulses}
\label{Reduce the number of data carriers with generalized pulses}

Since the sidelobes of the basic pulse, $P_k(f)$, decrease as we move away form the carrier frequency, the \ac{OOBE} of an OFDM signal is mainly due to the nearby carriers \cite{Weiss04bis}. This fact was exploited in \cite{Sahin11} by using shaping pulses with smoother transitions in the carriers close to the edges of the notched band than in distant ones. Similarly, the generalized pulses can be applied only to the set of carriers located in the vicinity of the bands where the \ac{OOBE} has to be lowered, $\mathcal{D}^h=\left\{d^h_1, \ldots, d^h_{|\mathcal{D}^h|}\right\}$. Distant carriers would use $p_k(n)$. There is no general rule for selecting $\mathcal{D}^h$, but those carriers whose \ac{PSD} exceeds the permitted mask when using $p_k(n)$ can be selected in a first attempt. The transmitted symbols are then given by 
\begin{equation}
\mathbf{x}_i=\mathbf{H}_{\mathcal{D}^h} \mathbf{s}_{\mathcal{D}^h}(i)+\mathbf{P}_{\left(\mathcal{D}-\mathcal{D}^h\right)} \mathbf{s}_{\left(\mathcal{D}-\mathcal{D}^h\right)}(i),
\label{ec.18bis_c}
\end{equation} 
where the first term on the \ac{RHS} corresponds to the carriers with generalized pulses while the second one to the carriers that use conventional pulses. The column vectors of $\mathbf{H}_{\mathcal{D}^h}$ are computed by solving (\ref{ec.10bisbisbis}) $\forall k \in \mathcal{D}^h$. 

This strategy reduces the number of products required to implement the terms due to the \ac{CC} and to the transition pulses in (\ref{ec.18}) to $|\mathcal{C}| \cdot |\mathcal{D}^h|$ and $2\beta|\mathcal{D}^h|$, respectively. 

\subsection{Reducing the set of \ac{CC} used in the generalized pulses}

This simplification is grounded in the above mentioned reasoning, but applied to the \ac{CC}. Accordingly, generalized pulses would only include \ac{CC} located in the edges of the closer notched bands. \ac{CC} in the edges of distant notches do not have to be included. As an example, consider the \ac{PSD} mask shown in Fig. \ref{Fig.1} (a). In this case, the generalized pulse of the data carrier in blue would only include the \ac{CC} in red. \ac{CC} in the vicinity of distant notches (not shown in the figure) would not be considered for this data carrier. 

Furthermore, only the \ac{CC} located in the same side of the notched band suffices in most cases. It has been verified that designating as \ac{CC} just two carriers located in the outer edge and one in the inner edge of the notched band provides the required \ac{OOBE} reduction in most cases. Referring to the example in Fig. \ref{Fig.1} (a), this means that the generalized pulse in blue would only employ the three \ac{CC} on the left edge of this notch: two inband and one outband. As it will be shown in Fig \ref{Fig.3}, this strategy lowers the \ac{OOBE} by almost 30 dB (with respect to the level of the conventional OFDM), even without using transition pulses.

This strategy makes the set of \ac{CC} to be different in each data carrier, $\mathcal{C}(k)=\left\{c_1(k), \ldots,c_{|\mathcal{C}(k)|}(k)\right\}$ for $k \in \mathcal{D}$. Hence, the generalized pulses are given by 
\begin{equation}
\mathbf{h}_k = \mathbf{p}_k + \Big[\mathbf{P}_{\mathcal{C}(k)} \;\; \mathbf{T} \Big] \left[ \begin{array}{c} \pmb{\alpha}_k \\ \pmb{\zeta}_k \end{array}\right]= \mathbf{p}_k + \pmb{\Pi} \, \pmb{\gamma}_k, \;\; k\in \mathcal{D},
\end{equation}
with $\pmb{\alpha}_k=\left[\alpha_k(1), \ldots, \alpha_k\left(|\mathcal{C}(k)|\right)\right]^T$ and where $\pmb{\gamma}_k$ is computed according to (\ref{ec.10bisbisbis}). The computational complexity of the term due to the CC reduces to $\sum_{k \in \mathcal{D}} |\mathcal{C}(k)|$. In practice, all the generalized pulses located close to an edge of a given notched band would use the same set of \ac{CC}.

\subsection{Employing transition pulses designed by windowing conventional pulses}
\label{Use transition pulses created by windowing complex exponentials}

The boundaries of the transition pulses are designed ad hoc for each data carrier. Since their length is usually much larger that the number of \ac{CC}, this significantly increases the complexity of the optimization problem and of the transmitter implementation. An intuitive approach to reduce the latter is to design the transition pulses using an expression formally equivalent to the one of $\mathbf{x}_i$ in (\ref{ec.1bisbis}), which can be efficiently implemented by means of the \ac{IDFT},
\begin{equation}
\mathbf{t}_k = \mathbf{U} \, \pmb{\Delta}_{N_{GI},\beta} \, \mathbf{W}_N^{\mathcal{Q}} \, \pmb{\lambda}_k \;\;k \in \mathcal{D},
\label{ec.19}
\end{equation}
where $\pmb{\lambda}_k=\left[\lambda_k(1), \ldots, \lambda_k(|\mathcal{Q}|)\right]^T$, with $\mathcal{Q}=\left\{q_1, \ldots, q_{|\mathcal{Q}|}\right\}$, $\mathbf{W}_N^{\mathcal{Q}}=\left[\mathbf{w}_N^{q_1},\ldots, \mathbf{w}_N^{q_{|\mathcal{Q}|}}\right]$ and $\pmb{\Delta}_{N_{GI},\beta}$ is the $L \times N$ matrix that performs the cyclic extension of the IDFT output, as defined in (\ref{ec.1bisbisbis}). Being $\mathbf{U}=\textrm{diag}\Big(u\left(0\right), \ldots, u\left(L-1\right)\Big)$, where $u(n) \in \mathbb{R}$ with non-zero values only in the edges of the generalized pulse, $n\in\left\{0,\ldots,\beta-1,L-\beta,\dots, L-1\right\}$. Its boundaries can have any convenient shape, e.g., a Hamming pulse of length $\beta$. In principle, $\mathcal{Q}=\mathcal{K}$ but, in practice, reducing the set size to $\mathcal{Q}=\mathcal{C}$ provides a good trade-off between \ac{OOBE} suppression and complexity, as it will be shown in Fig. \ref{Fig.5}. 

The generalized pulses can then be written as
\begin{equation}
\mathbf{h}_k = \mathbf{p}_k + \underbrace{\left[\mathbf{P}_{\mathcal{C}} \;\; \mathbf{T}_{w} \right]}_{\pmb{\Pi}^{\textrm{w}}}
\underbrace{
\left[
\begin{array}{c}
\pmb{\alpha}_k\\
\pmb{\lambda}_k
\end{array}
\right]
}_{\pmb{\gamma}^{\textrm{w}}_k}=\mathbf{p}_k + \pmb{\Pi}^{\textrm{w}} \, \pmb{\gamma}^{\textrm{w}}_k,
 \;\; k \in \mathcal{D},
\label{ec.20}
\end{equation}
where 
\begin{equation}
\mathbf{T}_w = \mathbf{U} \, \pmb{\Delta}_{N_{GI},\beta} \, \mathbf{W}_N^{\mathcal{Q}}.
\label{ec.20bis}
\end{equation}
Expression (\ref{ec.20}) is formally equivalent to (\ref{ec.9}). Hence, $\pmb{\gamma}^{\textrm{w}}_k$ can be computed from the minimization in (\ref{ec.10bisbisbis}) just by particularizing $E_{k,\mathcal{B}}$ with (\ref{ec.20}). The OFDM symbols can then be generated as                                            
\begin{equation}
\mathbf{x}_i = \mathbf{H}_{\mathcal{D}} \mathbf{s}_{\mathcal{D}}(i) =\mathbf{P}_{\mathcal{D}} \mathbf{s}_{\mathcal{D}}(i) + \mathbf{P}_{\mathcal{C}} \mathbf{A}_{\mathcal{D}} \mathbf{s}_{\mathcal{D}}(i) + \mathbf{T}_w \pmb{\Lambda}_{\mathcal{D}} \mathbf{s}_{\mathcal{D}}(i),
\label{ec.21}
\end{equation}
where $\pmb{\Lambda}_{\mathcal{D}}=\left[\hat{\pmb{\lambda}}_{d_1}, \ldots, \hat{\pmb{\lambda}}_{d_{|\mathcal{D}|}}\right]$. 

Hence, the additional complexity of this transmitter with respect to the conventional one is the second and the third terms on the \ac{RHS} of (\ref{ec.21}). The complexity of the second term is small and has been already discussed in Section \ref{Generalized method for spectral shaping-Transmitter implementation}. The third one can be obtained by means of the \ac{IDFT}, plus the product of the modulating symbols by $\pmb{\Lambda}_{\mathcal{D}}$ and the shaping process defined by the matrix $\mathbf{U}$, which involves $2\beta$ products per symbol. 

\subsection{Employing transition pulses with harmonically designed boundaries}
\label{Use harmonically designed transition pulses}
Designing the transition pulses according to the previous method is conceptually simple but still requires an $N$-point \ac{IDFT}. A simpler transmitter can be obtained by applying the series decomposition only to the boundaries of the transition pulses (which are the only non-zero samples) instead of to the whole pulse,
\begin{equation}
\mathbf{t}_k=\left[\begin{array}{c}\mathbf{W}_{\beta} \\ \mathbf{0}_{L-\beta,\beta} \end{array}\right] \, \pmb{\xi}_k^s + \left[\begin{array}{c}\mathbf{0}_{L-\beta,\beta} \\ \mathbf{W}_{\beta} \end{array}\right] \, \pmb{\xi}_k^e,
\label{ec.23}                                                                                             
\end{equation}
where  $\pmb{\xi}_k^s=\left[\xi_k^s(1), \ldots, \xi_k^s(\beta)\right]^T$ and $\pmb{\xi}_k^e=\left[\xi_k^e(1), \ldots, \xi_k^e(\beta)\right]^T$. 

It is worth noting that $\pmb{\xi}_k^s$ and $\pmb{\xi}_k^e$ can be seen as the $\beta$-point \ac{DFT} of the starting and ending boundaries of the transition pulse, respectively. In order to lower the \ac{PSD} in a given band, only the \ac{DFT} values corresponding to frequencies in the vicinity of this band need to be considered. Distant ones have almost no influence and can be set to zero. This notably simplifies the optimization problem and the implementation complexity, as it will be shown below. 
 
The generalized pulses can then be written as
\begin{equation}
\mathbf{h}_k=\mathbf{p}_k+\underbrace{\left[\mathbf{P}_{\mathcal{C}} \;\; \mathbf{T}_h\right]}_{\pmb{\Pi}^{\textrm{h}}} \underbrace{\left[ \begin{array}{c}\pmb{\alpha}_k \\ \pmb{\xi}_k^s \\ \pmb{\xi}_k^e \end{array}\right]}_{\pmb{\gamma}^{\textrm{h}}_k}=\mathbf{p}_k + \pmb{\Pi}^{\textrm{h}} \, \pmb{\gamma}^{\textrm{h}}_k, \;\; k \in \mathcal{D},
\label{ec.24}
\end{equation}
where $\mathbf{T}_h$ is given by
\begin{equation}
\mathbf{T}_h=\left[\begin{array}{cc}\mathbf{W}_{\beta} & \mathbf{0}_{L-\beta,\beta}\\ \mathbf{0}_{L-\beta,\beta} & \mathbf{W}_{\beta} \end{array}\right].
\label{ec.23bis}
\end{equation}
Since (\ref{ec.24}) is formally equivalent to (\ref{ec.9}), $\pmb{\gamma}^{\textrm{h}}_k$ is computed from (\ref{ec.10bisbisbis}) just by particularizing $E_{k,\mathcal{B}}$ with (\ref{ec.24}). The $i$-th OFDM symbol is then obtained as
\begin{equation}
\begin{aligned}
\mathbf{x}_i &= \mathbf{H}_{\mathcal{D}} \mathbf{s}_{\mathcal{D}}(i)=\mathbf{P}_{\mathcal{D}} \mathbf{s}_{\mathcal{D}}(i) + \mathbf{P}_{\mathcal{C}} \mathbf{A}_{\mathcal{D}} \mathbf{s}_{\mathcal{D}}(i) \\
&+ \left[\begin{array}{c}\mathbf{W}_{\beta}\\ \mathbf{0}_{L-\beta,\beta}\end{array}\right] \underbrace{\pmb{\Xi}_{\mathcal{D}}^s \mathbf{s}_{\mathcal{D}}(i)}_{\textstyle\mathbf{s}_{\mathcal{D}}^s(i)}
+ \left[\begin{array}{c}\mathbf{0}_{L-\beta,\beta}\\ \mathbf{W}_{\beta}\end{array}\right] \underbrace{\pmb{\Xi}_{\mathcal{D}}^e \mathbf{s}_{\mathcal{D}}(i)}_{\textstyle\mathbf{s}_{\mathcal{D}}^e(i)},
\end{aligned}
\label{ec.26}
\end{equation}
where $\pmb{\Xi}_{\mathcal{D}}^s=\left[\hat{\pmb{\xi}}_{d_1}^s, \ldots, \hat{\pmb{\xi}}_{d_{|\mathcal{D}|}}^s\right]$ and $\pmb{\Xi}_{\mathcal{D}}^e=\left[\hat{\pmb{\xi}}_{d_1}^e, \ldots, \hat{\pmb{\xi}}_{d_{|\mathcal{D}|}}^e\right]$. 

\vspace{0.1cm}
In principle, the computation of $\textstyle\mathbf{s}_{\mathcal{D}}^s(i)$ and $\textstyle\mathbf{s}_{\mathcal{D}}^e(i)$ requires multiplying the $\beta \times |\mathcal{D}|$ matrices $\pmb{\Xi}_{\mathcal{D}}^s$ and $\pmb{\Xi}_{\mathcal{D}}^e$ by the $|\mathcal{D}| \times 1$ vector $\mathbf{s}_{\mathcal{D}}(i)$. In practice, their computation is considerably simpler because of a twofold reason. The first is that generalized pulses are only used in a subset of the data carriers, $\mathcal{D}^h$, as justified in section \ref{Reduce the number of data carriers with generalized pulses}. The second is that $\pmb{\Xi}_{\mathcal{D}}^s$ and $\pmb{\Xi}_{\mathcal{D}}^e$ are sparse matrices because, as mentioned, the values of $\pmb{\xi}_k^s$ and $\pmb{\xi}_k^e$ corresponding to frequencies far away from the notched band are set to zero. 

We now concentrate in the third and the fourth terms on the \ac{RHS} of (\ref{ec.26}), since the first and the second ones have already been addressed. For convenience in notation, let us define $\tilde{\mathbf{x}}_i^s=\left[\tilde{x}_i^s(1), \ldots, \tilde{x}_i^s(\beta)\right]^T$ and $\tilde{\mathbf{x}}_i^e=\left[\tilde{x}_i^e(1), \ldots, \tilde{x}_i^e(\beta)\right]^T$ as the contribution of the transition pulses to the $\beta$ starting and ending samples of the $i$-th symbol,
\begin{equation}
\begin{array}{c}
\tilde{\mathbf{x}}_i^s=\mathbf{W}_{\beta} \, \mathbf{s}_{\mathcal{D}}^s(i),\\
\tilde{\mathbf{x}}_i^e=\mathbf{W}_{\beta} \, \mathbf{s}_{\mathcal{D}}^e(i).
\end{array}
\label{ec.27}
\end{equation}

As seen, each of them can be easily obtained by means of the $\beta$-point \ac{IDFT} of a linear combination of the input symbols. However, since the last $\beta$ samples of $\mathbf{x}_i$ overlap with the first $\beta$ samples of $\mathbf{x}_{i+1}$, a single $\beta$-point \ac{IDFT} per symbol is required, 
\begin{equation}
\tilde{\mathbf{x}}_i^e + \tilde{\mathbf{x}}_{i+1}^s = \mathbf{W}_{\beta} \bigg( \mathbf{s}_{\mathcal{D}}^e(i) + \mathbf{s}_{\mathcal{D}}^s(i+1) \bigg). 
\label{ec.28}
\end{equation}

Since $\beta$ is generally much smaller than $N$, the resulting transmitter is only moderately more complex than the conventional one and considerably much simpler than the one that uses the transition pulses defined in (\ref{ec.8}).

\section{Relation to previous methods}
\label{Relation to previous methods and novelties of the proposal}

\subsection{Relation to previous \ac{AIC} methods}
\label{Relation to previous AIC methods}
The method in \cite{Yamaguchi04} can be considered as a particular case of the generalized pulse in (\ref{ec.7}) when no transition pulses are used $(\pmb{t}_k=\mathbf{0}_{L,1})$, just one \ac{CC} at both outer edges of the notched band is employed and the optimization problem is solved without constraint. This work was extended in \cite{Brandes06} by incorporating a constraint to limit the peaks in the \ac{PSD}. However, this makes the optimization problem data-dependent, obliging to solve a \ac{LSQI}  problem in each OFDM symbol \cite{Golub96}. The same solution can be obtained by particularizing (\ref{ec.7}) with two \ac{CC} carriers at both inner edges of the band, $\pmb{t}_k=\mathbf{0}_{L,1}$ and by solving the optimization problem subject to the constraint $\left\|\pmb{\alpha}_k\right\|^2 \leq \epsilon$. Moreover, the framework presented in this paper also innovates the methods in \cite{Yamaguchi04} and \cite{Brandes06} by means of the reduced-complexity strategy described in section \ref{Reduce the number of data carriers with generalized pulses}.

\subsection{Relation to previous time-domain methods}
The method described in \cite{Sahin11} proposes the application of different windows to different carriers: one window with longer transition regions that is applied to carriers located at the edges of the band where the \ac{OOBE} has to be reduced, and another with shorter transition regions (or even a rectangular one) that is applied to the distant carriers. This can be easily obtained with the generalized pulse in (\ref{ec.7}) just by setting $\pmb{\alpha}_k=\mathbf{0}_{|\mathcal{C}|,1}$ and by constraining the transition pulses to $\pmb{t}_k=\tilde{\pmb{t}}$ for the inner carriers and $\pmb{t}_k=\check{\pmb{t}}$ for the carriers located in the edges, where $\tilde{\pmb{t}}$ and $\check{\pmb{t}}$ are predefined waveforms. 

In \cite{Mahmoud08}, a time-domain extension, referred to as \ac{AST}, is added to each OFDM symbol. This same concept can be implemented by particularizing the generalized pulse in (\ref{ec.9}) with no \ac{CC} $(\pmb{\alpha}_k=\mathbf{0}_{|\mathcal{C}|,1})$ and solving the optimization problem subject to the constraint $||\pmb{\zeta}_k||^2 \leq \epsilon$. Since the \ac{AST} is computed on a symbol basis, the solution obtained with the generalized pulse proposed is not exactly equal to the one in \cite{Mahmoud08}. However, both are grounded in the same principle and, while in the former a \ac{LSQI} optimization problem has to be solved in each OFDM symbol, the latter yields a data-independent design. 

\section{Numerical results}
\label{Results}


In the following subsections an OFDM system like the one defined in ITU-T Rec. G.9964 is considered \cite{G.9964}. It uses $N=4096$, $N_{GI}=1024$, $\beta=512$ and a sampling frequency of 100 MHz. The $\beta$ samples at both ends of $g(n)$ are always shaped using a \ac{RC} window when the generalized pulses are employed, unless otherwise stated. 

\subsection{Assessment of the spectral shaping with generalized pulses}

The performance of the generalized pulses when used to reduce the \ac{OOBE} in the frequency band, $\mathcal{B}$, corresponding to the carrier indexes $\left\{0, \ldots, 1024, 3022, \ldots, 3026, 3072, \ldots, 4095\right\}$ is assessed. This allows evaluating the capacity to reduce the emissions both in a spectral hole (corresponding to carrier indexes $3022$ to $3026$) and in the sidebands (carriers $0$ to $1024$ and $3072$ to $4095$). The problem is solved subject to the constraint on the magnitude of the real and imaginary parts in (\ref{ec.12_a}). All the \ac{PSD}s shown in this section have been obtained analytically by particularizing (\ref{ec.2}) with $P_k(f)$ or $H_k(f)$, as appropriate. 

\begin{figure}[h!]
\centering
  \includegraphics[width=0.6\columnwidth]{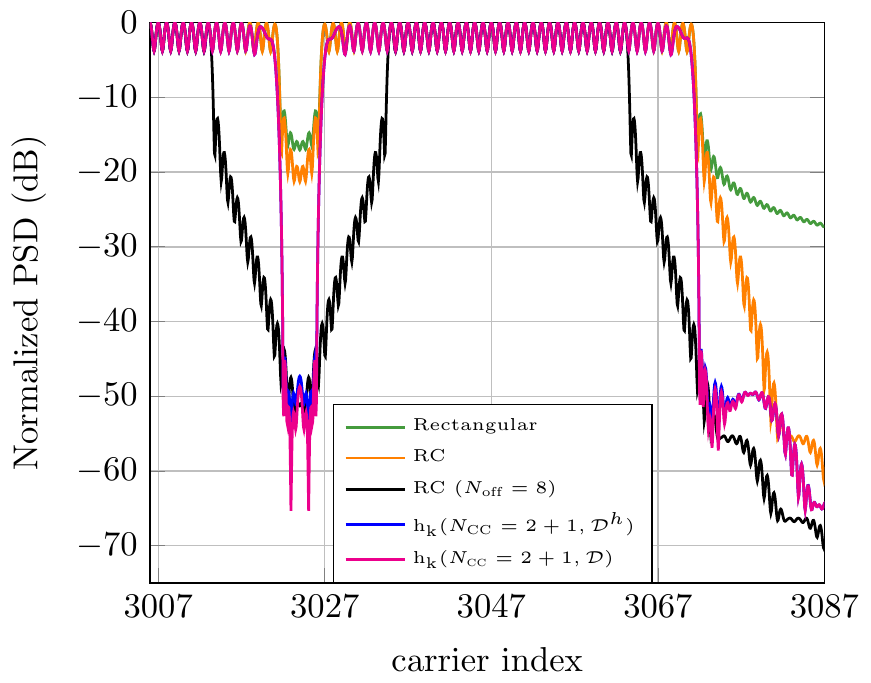}
  \caption{\ac{PSD} with pulse-shaping using an \ac{RC} window and with generalized pulses that only employ \ac{CC}. The former are labeled as \ac{RC} and latter as $\textrm{h}_{\textrm{k}}(N_{\protect\scalebox{0.75}{CC}}=N_{\protect\scalebox{0.75}{CC}}^D+N_{\protect\scalebox{0.75}{CC}}^B, \mathcal{D}^h)$ and $\textrm{h}_{\textrm{k}}(N_{\protect\scalebox{0.75}{CC}}=N_{\protect\scalebox{0.75}{CC}}^D+N_{\protect\scalebox{0.75}{CC}}^B, \mathcal{D})$, depending on the set of carriers that employ generalized carriers.}
\label{Fig.3}
\end{figure}

First, the performance achieved with generalized pulses that do only employ \ac{CC} is evaluated. Fig. \ref{Fig.3} depicts the normalized \ac{PSD} obtained when the \ac{OOBE} is reduced only by means of pulse-shaping using an \ac{RC} window. As a reference, the \ac{PSD} that results when $g(n)$ is a rectangular pulse with length $L=N+N_{GI}$ is also shown. As seen, the \ac{RC} windowing notably diminishes the \ac{OOBE} in the sideband: about 33 dB at carrier index $3087$. However, the reduction in the notched band is quite modest: 4 dB at most. 

Fig. \ref{Fig.3} also shows the performance achieved when using generalized pulses. Two cases are considered, one in which the generalized pulses are used in all data carriers and another in which they are only used in the reduced set $\mathcal{D}^h$. In the latter, the remaining carriers use standard \ac{RC} pulses. $\mathcal{D}^h$ includes the $N_D=9$ data carriers closer to the edges of the four bands where the \ac{OOBE} has to be reduced, hence, $|\mathcal{D}^h|=36$. All the generalized pulses use the same number of \ac{CC}, $|\mathcal{C}|=4N_{\scalebox{0.75}{CC}}$, where $N_{\scalebox{0.75}{CC}}$ denotes the number of \ac{CC} at each edge of the four notched bands. $N_{\scalebox{0.75}{CC}}=N_{\scalebox{0.75}{CC}}^D+N_{\scalebox{0.75}{CC}}^B$, with $N_{\scalebox{0.75}{CC}}^D$ being the number of \ac{CC} data carriers now used as \ac{CC} (inband \ac{CC}), and $N_{\scalebox{0.75}{CC}}^B$ is the number of \ac{CC} placed inside the notched band (outband \ac{CC}). For the sake of clarity, the resulting cases are designated as $h_k(N_{\scalebox{0.75}{CC}}=N_{\scalebox{0.75}{CC}}^D+N_{\scalebox{0.75}{CC}}^B, \mathcal{D})$ and $h_k(N_{\scalebox{0.75}{CC}}=N_{\scalebox{0.75}{CC}}^D+N_{\scalebox{0.75}{CC}}^B, \mathcal{D}^h)$, depending on the set of carriers where the generalized pulses are employed. It is interesting to recall that the case in which the generalized pulses are used in the set $\mathcal{D}$ is equivalent to the method proposed in \cite{Yamaguchi04}, except for the use of three \ac{CC} instead of one, and to the one in \cite{Brandes06}, except for the constraint of the optimization problem and the use $N_{\scalebox{0.75}{CC}}^B$ outband \ac{CC}. 

As seen, the difference between the reduced-complexity configuration with $\mathcal{D}^h$ and the one with $\mathcal{D}$ is negligible. In both cases, the \ac{PSD} level in the notched band is decreased by about 25 dB with respect to the case where only the \ac{RC} windowing is employed. The magnitude of this reduction is better quantified in terms of the number of carriers that have to be nulled to achieve the same \ac{PSD} level when the pulse-shaping with \ac{RC} is employed. As shown in Fig. \ref{Fig.3}, $N_{\textrm{off}}=8$ data carriers are required to this end. In the sideband, the \ac{PSD} obtained with the generalized pulses decays very abruptly at the beginning, but at high frequencies it tends to the \ac{PSD} given by the pulse-shaping with \ac{RC}. 

\begin{figure}[h!]
\centering
  \includegraphics[width=0.6\columnwidth]{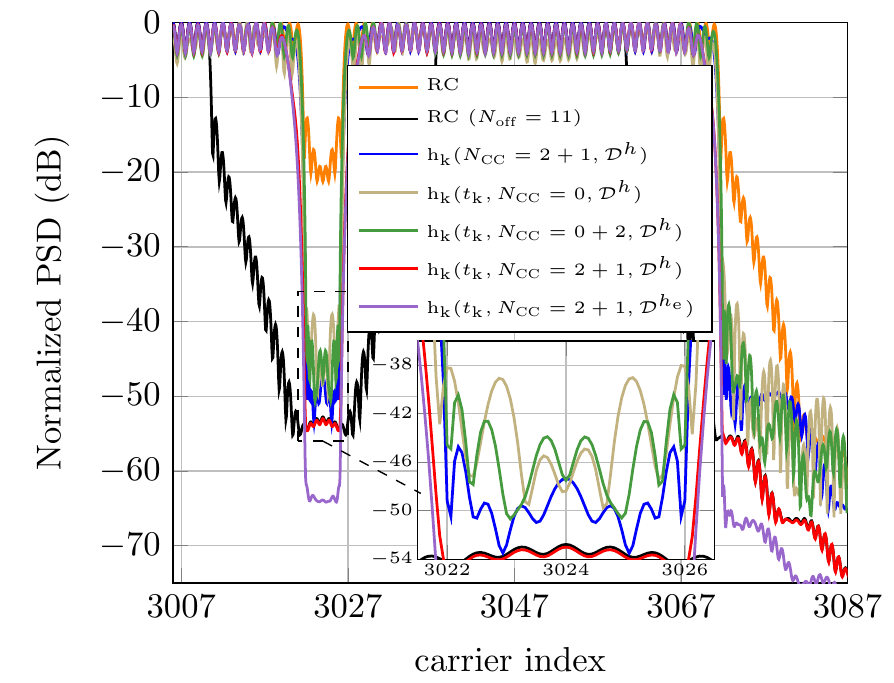}
	\caption{\ac{PSD} with pulse-shaping using an \ac{RC} window and with generalized pulses. The former are denoted as \ac{RC} and latter as $\textrm{h}_{\textrm{k}}(N_{\protect\scalebox{0.75}{CC}}, \mathcal{D}^h)$ when only \ac{CC} are employed. Curves labeled as $\textrm{h}_{\textrm{k}}(t_k, N_{\protect\scalebox{0.75}{CC}}, \mathcal{D}^h)$ and $\textrm{h}_{\textrm{k}}(t_k, N_{\protect\scalebox{0.75}{CC}},\mathcal{D}^{h_{\protect\scalebox{0.75}{e}}})$ correspond to the cases where transition pulses and \ac{CC} are used in the sets $\mathcal{D}^h$ and $\mathcal{D}^{h_{\protect\scalebox{0.75}{e}}}$, respectively.} 
\label{Fig.4}
\end{figure}

The \ac{OOBE} reduction obtained when using transition pulses, with and without \ac{CC}, is now assessed. Fig. \ref{Fig.4} depicts the \ac{PSD}s obtained with different generalized pulses and with pulse-shaping using an \ac{RC} window, which serves as reference. The case where transition pulses are used is denoted as $h_k(t_k, N_{\scalebox{0.75}{CC}}, \mathcal{D}^h)$. The comparison of the cases $h_k(N_{\scalebox{0.75}{CC}}=2+1, \mathcal{D}^h)$ and $h_k(t_k, N_{\scalebox{0.75}{CC}}=0, \mathcal{D}^h)$ reveals that the \ac{CC} achieve larger \ac{OOBE} reductions than the transition pulses. As expected, the combined use of both cancellation terms considerably improves the performance and the number of carriers that have to be nulled to achieve the same \ac{PSD} level with an \ac{RC} windowing increases from $N_{\textrm{off}}=8$ to $N_{\textrm{off}}=11$. However, it is interesting to notice that the case $h_k(t_k, N_{\scalebox{0.75}{CC}}=0+2, \mathcal{D}^h)$ achieves a significant \ac{OOBE} reduction in the band corresponding to carriers 3022 to 3026 and avoids the bit-rate loss caused by using data carriers as \ac{CC} (inband \ac{CC}). 

Fig. \ref{Fig.4} also depicts the \ac{PSD} of the case $h_k(t_k, N_{\scalebox{0.75}{CC}}=2+1,\mathcal{D}^{h_{\scalebox{0.75}{e}}})$, where $\mathcal{D}^{h_{\scalebox{0.75}{e}}}$ is an extended set that comprises the $N_D=15$ data carriers closer to each edge of the four bands where the \ac{OOBE} has to be reduced, therefore, $|\mathcal{D}^{h_{\scalebox{0.75}{e}}}|=60$. Incrementing $N_D$ yielded no performance improvement when the \ac{CC} are the only cancelling term, as shown in Fig. \ref{Fig.3}. However, the addition of a new degree of freedom to the problem, in the form of transition pulses, augments the \ac{OOBE} reduction capacity of the generalized pulses. Consequently, the distance at which the latter are able to lower the \ac{PSD} also increases, as seen by comparing $h_k(t_k, N_{\scalebox{0.75}{CC}}=2+1,\mathcal{D}^{h_{\scalebox{0.75}{e}}})$ to  $h_k(t_k, N_{\scalebox{0.75}{CC}}=2+1, \mathcal{D}^h)$. Finally, it must be emphasized that $h_k(N_{\scalebox{0.75}{CC}}=2+1,\mathcal{D}^h)$ indeed combines \ac{AIC} and time-domain methods because the basic pulse, $p_k(n)$, uses an \ac{RC} shaping window. However, since only the frequency-domain cancellation term is optimized, it performs much worse than $h_k(t_k, N_{\scalebox{0.75}{CC}}=2+1, \mathcal{D}^h)$, where a joint optimization of the frequency and time-domain cancellation terms is accomplished. 

The OOBE achieved when the generalized pulses are used in the set $\mathcal{D}^h$ suffices for most applications. Hence, this set is employed from now on and the focus is put on the reduced-complexity strategies proposed in sections \ref{Use transition pulses created by windowing complex exponentials} and \ref{Use harmonically designed transition pulses} to design the transition pulses.

\begin{figure}[h!]
\centering
  \includegraphics[width=0.6\columnwidth]{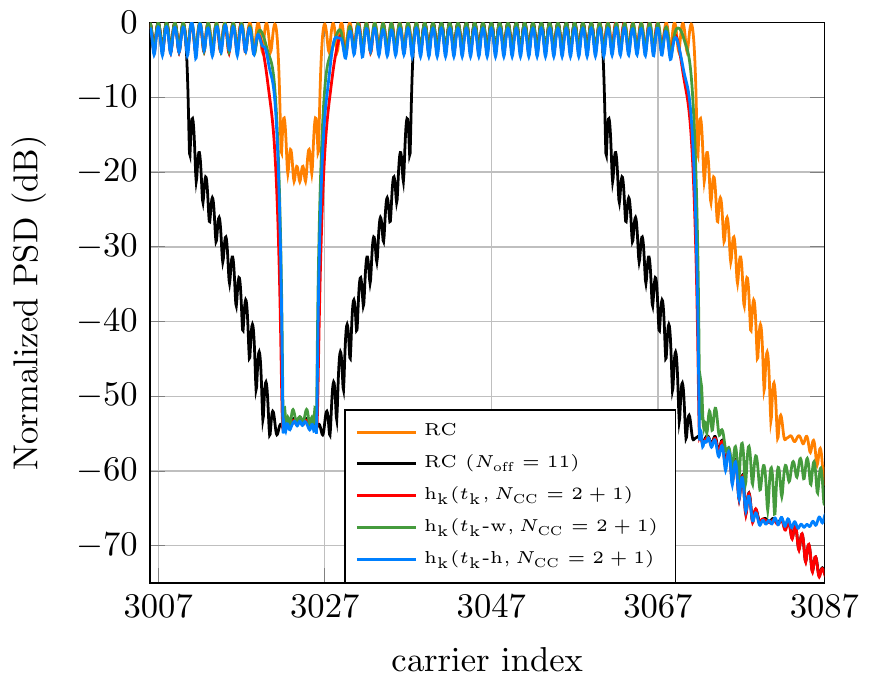}
  \caption{\ac{PSD} with pulse-shaping using an \ac{RC} window and with generalized pulses whose transition pulses are obtained by: the general case in (\ref{ec.8}),  $\textrm{h}_{\textrm{k}}(t_k, N_{\protect\scalebox{0.75}{CC}})$; windowing conventional pulses, $\textrm{h}_{\textrm{k}}(t_k \textrm{-w}, N_{\protect\scalebox{0.75}{CC}})$; harmonically designed boundaries, $\textrm{h}_{\textrm{k}}(t_k \textrm{-h}, N_{\protect\scalebox{0.75}{CC}})$.} 
\label{Fig.5}
\end{figure}

Fig. \ref{Fig.5} shows the \ac{PSD} obtained with pulse-shaping using an \ac{RC} window and with generalized pulses that employ \ac{CC} and transition pulses designed according to different criteria. The general case in which transition pulses are given by (\ref{ec.9}) is denoted as $\textrm{h}_{\textrm{k}}(t_k, N_{\scalebox{0.75}{CC}})$, while the one in which transition pulses are obtained by windowing conventional pulses, as proposed in section \ref{Use transition pulses created by windowing complex exponentials}, is designated as $\textrm{h}_{\textrm{k}}(t_k \textrm{-w}, N_{\scalebox{0.75}{CC}})$. The latter have been generated according to (\ref{ec.19}) using $\mathcal{Q}=\mathcal{C}$. The \ac{PSD} obtained when transition pulses have harmonically designed boundaries, as described in \ref{Use harmonically designed transition pulses}, is referred to as $\textrm{h}_{\textrm{k}}(t_k \textrm{-h}, N_{\scalebox{0.75}{CC}})$. Only three non-zero terms have been employed in the harmonic series decomposition of the boundaries in (\ref{ec.23}): the ones with discrete-time frequencies closer to the edges of the notched bands. As seen, performance degradation due to constraining the waveform of the transition pulses is negligible, in particular when using harmonically designed boundaries.

\subsection{Comparison with previous methods}
\begin{figure}[h!]
\centering
  \includegraphics[width=0.6\columnwidth]{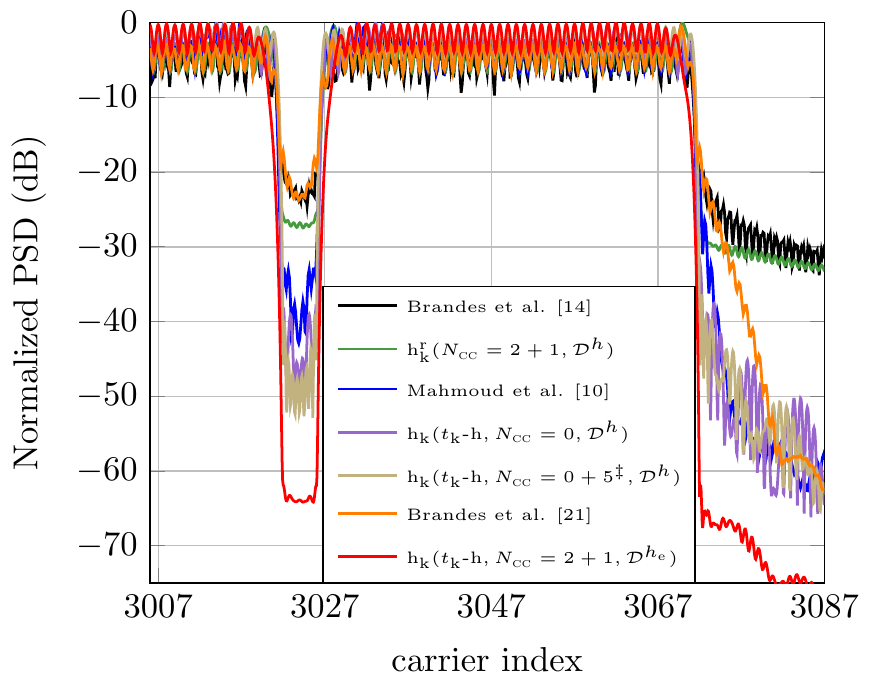}
	\caption{\ac{PSD} obtained with different configurations of the proposed method and others taken from the literature. The case $\textrm{h}^{\textrm{r}}_{\textrm{k}}(N_{\protect\scalebox{0.75}{CC}}=2+1, \mathcal{D}^h)$ does only employ \ac{CC} and a rectangular shaping pulse. $\textrm{h}_{\textrm{k}}(t_k \textrm{-h}, N_{\protect\scalebox{0.75}{CC}}=0, \mathcal{D}^h)$ denotes the case where only transition pulses are employed. While $\textrm{h}_{\textrm{k}}(t_k \textrm{-h}, N_{\protect\scalebox{0.75}{CC}}=0+5^{\ddagger}, \mathcal{D}^h)$ and $\textrm{h}_{\textrm{k}}(t_k \textrm{-h}, N_{\protect\scalebox{0.75}{CC}}=2+1, \mathcal{D}^{h_{_{\protect\scalebox{0.75}{e}}}})$ combine both \ac{CC} and transition pulses, the former only uses outband \ac{CC}.} 
\label{Fig.6b}
\end{figure}

Fig. \ref{Fig.6b} assesses the \ac{OOBE} reduction achieved by the proposed method and others taken from the literature. The following techniques are appraised: the \ac{AIC} one by Brandes et al. \cite{Brandes06}, the time-domain method by Mahmoud et al. \cite{Mahmoud08} and the combination of \ac{AIC} and time-domain by Brandes et al. \cite{Brandes10}. 

The \ac{PSD} that results with the proposed method is analytically computed as in (\ref{ec.2}). In the remaining cases, the Welch's averaged periodogram method with a 16384-sample Hanning window and 4096-sample overlap is applied to an \ac{OFDM} signal consisting of 2000 QPSK modulated symbols. 


The generalized pulses are denoted as in the previous subsection, except for the case whose label begin as $\textrm{h}^{\textrm{r}}_{\textrm{k}}$, in which the superscript indicates the use of a rectangular shaping pulse, $g(n)$, to allow a fair comparison with \cite{Brandes06}. Hence, $\textrm{h}^{\textrm{r}}_{\textrm{k}}(N_{\scalebox{0.75}{CC}}=2+1, \mathcal{D}^h)$ is conceptually equivalent to \cite{Brandes06} although, as mentioned in subsection \ref{Relation to previous AIC methods}, the former uses a different constraint in the optimization problem and an additional outband \ac{CC}. As seen, the proposed method creates a deeper notch in the passband. 

The case $\textrm{h}_{\textrm{k}}(t_k \textrm{-h}, N_{\scalebox{0.75}{CC}}=0, \mathcal{D}^h)$ can be classified as a time-domain method. Nevertheless, it uses a different optimization criterion to the one in \cite{Mahmoud08}, as can be seen by comparing (\ref{ec.10bisbisbis}) to \cite[Eq. (5)]{Mahmoud08}. It can be observed that the latter performs slightly better in the sideband but the former creates a deeper notch. The proposed method can further lower the \ac{OOBE} without data rate penalty by using outband \ac{CC}. This is illustrated by the case $\textrm{h}_{\textrm{k}}(t_k \textrm{-h}, N_{\scalebox{0.75}{CC}}=0+5^{\ddagger}, \mathcal{D}^h)$, where the use as outband \ac{CC} of the 5 carriers with indexes 3022 to 3026 provides additional reductions of the \ac{OOBE} in this band. 

The AIC and time-domain combination in \cite{Brandes10} does only optimize the frequency-domain term. Its performance in the sideband approaches to the one of \cite{Mahmoud08}, but in the notched band it is almost equal to the one of \cite{Brandes06} because the optimization of the \ac{CC} is unaware of the subsequent pulse-shaping. As seen, $\textrm{h}_{\textrm{k}}(t_k \textrm{-h}, N_{\scalebox{0.75}{CC}}=2+1, \mathcal{D}^{h_{_{\scalebox{0.75}{e}}}})$ significantly outperforms both \cite{Mahmoud08} and \cite{Brandes10}, since it jointly optimizes the frequency and time-domain cancelling terms. 

\subsection{Performance example in an actual scenario}

This section assesses the capability of the generalized pulses to comply with a stringent \ac{PSD} mask. To this end, the in-home broadband \ac{PLC} scenario has been selected. In Europe, these system have to comply with the EN 50561-1. It defines 20 permanently excluded subbands within the range 1.8-30 MHz \cite{EN50561-1}. The \ac{PSD} level must drop by at least 43 dB in the notched subbands located below 5 MHz and by at least 39 dB in the ones located in the 5-30 MHz range \cite{EN55022}. Additionally, the ITU-T Rec. G.9964 imposes that the \ac{OOBE} below 1.8 MHz and above 30 MHz must be at least 30 dB lower than in the used band \cite{G.9964}. 
To further illustrate its the severity, it is worth mentioning that the width of the narrower notch is 50 kHz and that the width of the narrower data subband between two consecutive notches is 300 kHz. Expressed in terms of the carrier spacing, $\Delta f=100/4096 \approx 24.41$ kHz, their equivalent widths are 2 and 12 carriers, respectively. 


Applying the time-domain method in \cite{Mahmoud08} to the considered problem obliges to solve an \ac{LSQI} optimization problem with $\beta=512$ unknowns for each \ac{OFDM} symbol. The complexity of the \ac{AIC} methods in \cite{Brandes06} and \cite{Brandes10} is lower because they can profit from the fact that the \ac{OOBE} is mainly due to the nearest carriers, which allows dividing the large optimization problem into smaller ones. However, in the considered scenario, these smaller problems can still have a considerable number of unknowns to be solved for each \ac{OFDM} symbol. On the contrary, the optimization of the generalized pulses proposed in this work is accomplished offline and on an individual basis, since each generalized pulse embeds all the terms required to reduce its spectrum in the notched subbands. 

Two different configurations of generalized pulses are evaluated. The first one only uses \ac{CC} and is denoted as $\textrm{h}_{\textrm{k}}(N_{\scalebox{0.75}{CC}})$. The number of \ac{CC} at each edge of the notched subbands is always $N_{\scalebox{0.75}{CC}}=N_{\scalebox{0.75}{CC}}^D+N_{\scalebox{0.75}{CC}}^B=2+1$. The generalized pulses are only used in the $N_D$ data carriers closer to the edges of the notched subbands, where $N_D$ varies between 4 and 9. The second scheme uses both \ac{CC} and transition pulses with harmonically designed boundaries. It is referred to as $\textrm{h}_{\textrm{k}}(t_k\textrm{-h}, N_{\scalebox{0.75}{CC}})$. The use of the transition pulses allows reducing the number of data carriers used as \ac{CC} to $N_{\scalebox{0.75}{CC}}^D=1$. The number of outband \ac{CC} is generally set to $N_{\scalebox{0.75}{CC}}^B=1$, but in the most severe cases $N_{\scalebox{0.75}{CC}}^B=2$ is employed. The number of terms in the harmonic series decomposition of $\mathbf{t}_k$ in (\ref{ec.23}) is fixed to 5, except for some cases where 3 terms suffices. $N_D$ varies from 4 to 9.  

Table \ref{Table.2} compares the performance and the implementation complexity of the aforementioned configurations to the one of the pulse-shaping using an \ac{RC} window. The computational complexity of the latter transmitter is used as a reference. As seen, almost 28\% of the data carriers have to be disabled when using \ac{RC} windowing. The use of $\textrm{h}_{\textrm{k}}(N_{\scalebox{0.75}{CC}})$ notably reduces the performance loss, which in this case is due to the use of data carriers as \ac{CC}, with a negligible increment in the computational cost. The data carrier loss can be almost halved by using $\textrm{h}_{\textrm{k}}(t_k\textrm{-h}, N_{\scalebox{0.75}{CC}})$. Despite its cost is about 16\% larger than the one of the \ac{RC} windowing, this scheme is still simpler than multicarrier modulations with non-rectangular pulses.

The \ac{PAPR} is an important concern when dealing with OFDM signals \cite{Lim09}. It is well-known that time-domain spectral shaping techniques have nearly no influence on it, while \ac{AIC} methods increase it \cite{Mahmoud08}, \cite{Sahin11},\cite{Brandes06}, \cite{vandeBeek08}. Table \ref{Table.2} shows the \ac{PAPR} increment for a $10^{-3}$ clipping rate of the considered schemes with respect to the one that uses pulse-shaping with an \ac{RC} window, which is taken as a reference. Unsurprisingly, all values are quite modest, since most data carriers use conventional pulses and the generalized ones are employed only in a reduced subset. As expected, $\textrm{h}_{\textrm{k}}(N_{\scalebox{0.75}{CC}})$ gives the largest value. When transition pulses are incorporated, the influence of the \ac{CC} decreases and so does the \ac{PAPR}. In fact, it is slightly lower than the reference. The reason is that transition pulses only modify the boundaries of the basic pulse, whose amplitudes are smaller than in the inner part of the symbol because of the \ac{RC} windowing. As a result, the instantaneous amplitudes in the inner part, which dominate the \ac{PAPR}, are reduced at the cost of increasing the amplitudes in the edges. 


\begin{table}[h]
\renewcommand{\arraystretch}{1.3}
\setlength\tabcolsep{3pt}
\caption{Performance and complexity of different spectral shaping methods when used to comply with the EN 50561-1 and the ITU-T Rec. G.9964}
\label{Table.2}
\centering
\begin{tabular}{cccc}
\cline{2-4}
&\multicolumn{3}{c}{\textbf{Method}} \\
\cline{2-4}
& \renewcommand{\arraystretch}{0.8}\begin{tabular}{c}Pulse-shaping with \\ \ac{RC} window \end{tabular} & $\textrm{h}_{\textrm{k}}(N_{\scalebox{0.75}{CC}}=2+1)$ & $\textrm{h}_{\textrm{k}}(t_k\textrm{-h}, N_{\scalebox{0.75}{CC}})$  \\
\arrayrulecolor{black}\hline
\textbf{Data carriers loss (\%)} & 27.81 & 7.85 & 3.93 \\
\textbf{Products/symbol increment (\%)} & 0 & 2.60 & 15.58\\
\textbf{PAPR increment (dB)} & 0 & 0.27 & -0.05\\
\arrayrulecolor{black}\hline
\end{tabular}
\end{table}

\vspace{-0.5cm}
\section{Conclusion}
\label{Conclusion}
This work has proposed a method to shape the spectrum of \ac{OFDM} signals without affecting the receiver operation. It is grounded on the use of a generalized pulse that consists of the one employed in conventional OFDM systems plus a canceling term that reduces the spectrum of the former in the desired frequency bands. The canceling term has two components: a set of \ac{CC} and a transition pulse that only modifies the boundaries of the generalized pulse.

The presented approach generalizes and unifies previous \ac{AIC} and time-domain \ac{OOBE} suppression strategies. Moreover, while in most of the latter an optimization problem has to be solved for each OFDM symbol, the proposed technique results in a data-independent solution that can be performed offline, which notably simplifies the transmitter implementation, and allows the analytical calculation of the resulting \ac{PSD}. The latter can be further improved with negligible impact on the performance by designing the transition pulses according to two proposed methods that yield efficient \ac{IDFT}-based implementations. Hence, the proposed framework offers a wide range of alternatives with different trade-offs between implementation complexity and \ac{OOBE} reduction. 

It has been shown that a generalized pulse with only three \ac{CC} at each edge of the band where the \ac{PSD} has to be lowered (two in the outer edge and one in the inner edge) lowers the \ac{OOBE} achieved with pulse-shaping using an \ac{RC} window by about 25 dB. Increasing the number of \ac{CC} provides a negligible improvement. Transition pulses can further reduce the \ac{OOBE} by approximately 20 dB.  

In such a strict scenario as the one imposed to indoor broadband \ac{PLC} systems in Europe, with 20 permanently excluded subbands in the band 1.8-30 MHz, the proposed framework provides a series of solutions whose data carrier loss range from about 8\% (with complexity increment lower than 3\%) to less than 4\% (with complexity increment about 16\%). As a reference, almost 28\% of the data carriers have to be nulled when a conventional pulse-shaping with an \ac{RC} window is employed in this scenario.


\bibliographystyle{IEEEtran}
\bibliography{../biblio}

%

%
%
%
%




\end{document}